# Topological Magnetic-Spin Textures in Two-Dimensional Van Der Waals $Cr_2Ge_2Te_6$

Myung-Geun Han[1*], Joseph A. Garlow[1,2*], Yu Liu[1], Huiqin Zhang[3], Jun Li[1], Donald DiMarzio[4], Mark W. Knight[4], Cedomir Petrovic[1], Deep Jariwala[3], and Yimei Zhu[1,2]

[1] Condensed Matter Physics and Materials Science, Brookhaven National Laboratory, Upton, NY 11973, USA

[2] Department of Materials Science and Chemical Engineering, Stony Brook University, Stony Brook, NY, 11794, USA

[3] Department of Electrical and Systems Engineering, University of Pennsylvania, Philadelphia, PA 19014 USA

[3] NG Next, Northrop Grumman Corporation, Redondo Beach, CA 90278 USA

Correspondence to: mghan@bnl.gov

*These authors contributed equally.

**Abstract**

Two-dimensional (2D) van der Waals (vdW) materials show a range of profound physical properties that can be tailored through their incorporation in heterostructures and manipulated with external forces[1–5]. The recent discovery of long-range ferromagnetic order down to atomic layers provides an additional degree of freedom in engineering 2D materials and their heterostructure devices for spintronics, valleytronics and magnetic tunnel junction switches[6–9]. Here, using direct imaging by cryo-Lorentz transmission electron microscopy we show that topologically nontrivial magnetic-spin states, skyrmionic bubbles, can be realized in exfoliated insulating 2D vdW $Cr_2Ge_2Te_6$. Due to the competition between dipolar interactions and uniaxial magnetic anisotropy, hexagonally-packed nanoscale bubble lattices emerge by field cooling with magnetic field applied along the out-of-plane direction. Despite a range of topological spin textures in stripe domains arising due to pair formation and annihilation of Bloch lines, bubble lattices with single chirality are prevalent. Our observation of topologically-nontrivial homochiral skyrmionic bubbles in exfoliated vdW materials provides a new avenue for novel quantum states in atomically-thin insulators for magneto-electronic and quantum devices.

**Keywords:** van der Waals, topological, two-dimensional, ferromagnetism, Skyrmions, spintronics, Lorentz microscopy, Dzyaloshinskii-Moriya.

The isolation of a diverse range of atomically-thin two-dimensional (2D) layers from bulk van der Waals (vdW) materials has enabled new avenues in materials science, condensed matter physics and device engineering. Despite observation of all major electronic classes in 2D vdW layers (metal, insulators and semiconductors), the observation of ferromagnetic phases was missing until its discovery in Cr halides and chalcogenides in 2017[10,11]. Following this discovery, 2D magnets with long-range order have been investigated extensively for their fundamental spin physics in low dimensions and for their potential applications to heterostructures with other 2D materials for the advent of widely tunable properties[10–13]. Utilizing long-range ferromagnetic order, anomalous quantum Hall effects and spin filtered tunneling across 2D vdW heterostructures have been demonstrated[14–16].

Beyond simple ferromagnetism, skyrmionic spin textures are an exciting avenue for design and realization of topologically-driven, strongly-correlated quantum states in exfoliated 2D vdWs materials and their heterostructures for driving quantum coherence in devices. Such skyrmionic textures have been reported in chiral magnets, such as B20 metals and insulating $Cu_2OSeO_3$ in which the antisymmetric super exchange interaction (or Dzyaloshinskii-Moriya interaction, DMI) competes with dominating ferromagnetic exchange interaction[17–19]. In this case, the chirality (or handedness) is single-valued for a given crystal, as the sign of DMI vector is determined by underlying crystal structures. The individual skyrmion size and hexagonal skyrmion-lattice parameter are independent of magnetic field and fixed by the ratio between the magnitudes of ferromagnetic exchange interaction and the DMI. Particle-like skyrmions are of interest for future spintronic applications as they can be easily manipulated by spin current and strain.

Topologically-nontrivial spin textures have also been reported in centrosymmetric magnets due to a competition between the magnetic dipolar interaction and magnetocrystalline anisotropy[20,21]. For ferromagnetic thin films with an anisotropy ratio Q (= $K_u/K_d$) >> 1 and an easy axis perpendicular to the film plane, stripe domains spontaneously develop even for arbitrarily small thickness where $K_u$ is defined as the uniaxial anisotropy coefficient while $K_d$ is the stray field energy coefficient. Cylindrical-shape bubbles emerge before the field-induced saturation as a metastable state, similar to emergence of skyrmions in the vicinity of ferromagnetically poled states in chiral magnets. The lowest energy state of a bubble is characterized by a cylindrical Bloch wall with a single rotation sense. However, due to the lack of symmetry breaking DMI, internal spin structures, such as Bloch lines and points, can occur within the Bloch walls of stripe domains and bubbles, leading to a variety of topological configurations that are classified with their distinctive topological charges[21,22]; type-I without Bloch lines or points, and type-II with Bloch lines or points[23]. These Bloch lines and points dynamically form and annihilate due to their small formation energy and may provide an avenue to easily tune the topological charge in device architectures. Type-I bubbles and Bloch-type skyrmions in chiral magnets are indistinguishable in their topology as both carry the same topological charge +1 or -1 when mapped into the order parameter space, while type-II bubble's topological charge is 0. As a result, their spin textures manifest similarly during spin-charge transport, leading to topological Hall effects. For the type-I skyrmionic bubbles, the chirality is randomly chosen, and the size of bubbles is inversely scaled with external magnetic fields. Furthermore, the bubbles in a lattice are arranged in a hexagonally close-packed way and easily manipulated by electric current.

In this study, we show that topologically nontrivial spin textures, Bloch line pairs and skyrmionic bubbles, can be realized in exfoliated 2D vdW $Cr_2Ge_2Te_6$. Due to strong magnetic anisotropy with an out-of-plane easy axis, the ground-state stripe domains in exfoliated flakes turn into skyrmionic bubbles with external magnetic field applied perpendicular to the *c*-axis. Unlike conventional magnetic bubbles, the skyrmionic bubbles in CGT ($Cr_2Ge_2Te_6$) are homochiral (or single-chiral). Particle-like topological spin textures in insulating CGT can provide avenues to practically realize various theoretically proposed applications such as ultrafast and ultralow-power digital as well as synaptic, neuromorphic logic[24–26]. In addition, due to the extended nature of these spin textures when incorporated into 2D heterostructures with materials that host other quantum states such as single photon emitters[27] could provide an avenue for coherent coupling over long ranges for quantum information devices.

2D vdW CGT is a ferromagnetic insulator with an indirect band gap of ~ 0.74 eV, strong uniaxial anisotropy and an easy axis perpendicular to the *ab* planes – rendering it a promising platform to realize skyrmionic bubbles and Bloch lines/points in 2D vdW materials[11,28,29]. The space group of CGT is centrosymmetric $R\bar{3}$. The Te atoms are hexagonally packed in ABAB stacking sequence along the *c*-axis in which each AB layer forms a network of edge-shared Te octahedra. We examined crystal structure of CGT crystals by atomic resolution high-angle annular dark-field (HAADF) scanning transmission electron microscopy (STEM), as shown in Fig. 1. TEM samples from CGT crystals grown by chemical vapor method are prepared by mechanical exfoliation and dry transfer[30] for imaging along the [001] direction and focused ion beam (FIB) milling with 2 keV $Ga^+$ ion for imaging along other directions. The ABAB stacking of Te atoms are clearly observed in Figs 1c and e. The magnetic $Cr^{3+}$ ions (S = 3/2) and Ge pairs occupy Te-octahedral sites in a 1:2 ratio in each AB layer that is separated by a weak vdW bonding, as seen in Figs. 1e and f. In [211] projection imaging (Fig. 1c), we clearly see the pairs of Ge that occupy Te octahedral sites, as shown with green lines.

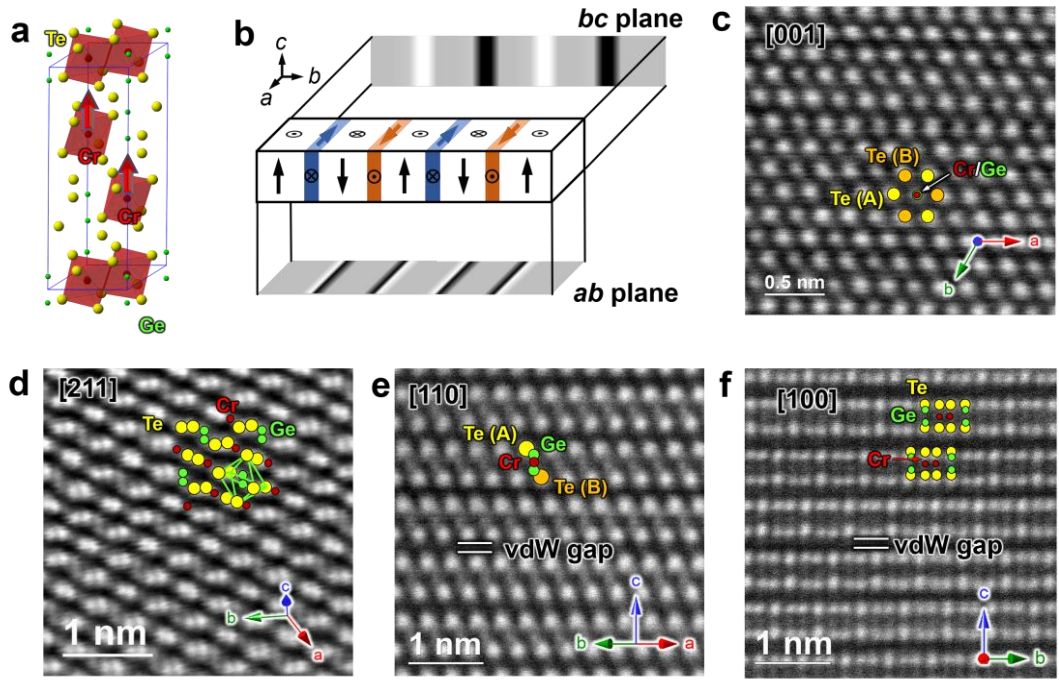

**Fig. 1| Atomic structure of 2D vdW Cr$_2$Ge$_2$Te$_6$ and stripe domains due to uniaxial anisotropy. a**, Cr$_2$Ge$_2$Te$_6$ unit cell showing CrTe$_6$ octahedra and Ge pairs. Red arrows depict the spin moments for Cr atoms along the *c*-axis. **b**, Schematic of stripe domains due to uniaxial anisotropy and their LTEM contrasts projected on *bc* and *ab* planes. Stripe domains with antiparallel spins are separated by Bloch walls. **c-f**, HAADF STEM images along the [001] (**c**), [211] (**d**), [110] (**e**), and [100] (**f**) axes show atomic structures of Cr$_2$Ge$_2$Te$_6$ with the space group $R\bar{3}$. In **d**, Te octahedra surrounding Ge pair is indicated with green lines.

Ferromagnetism arises from the super exchange interaction between nearest-neighbor $Cr^{3+}$ ions linked by Te ligands through nearly 90° angles, which shows a monoaxial anisotropy with the easy axis along the *c*-axis[31]. The ferromagnetically ordered Cr spins aligned along the *c*-axis are found to be stable down to double atomic-layers, as recently demonstrated[11]. Similar to bubble domains in thin films with an anisotropy ratio Q >> 1 and an easy axis perpendicular to the film plane, stripe domains separated by Bloch walls spontaneously develop without external magnetic field, as schematically shown in Fig. 1b. The incident electrons in transmission electron microscope passing through the sample in TEM are subject to phase shift induced by magnetization perpendicular to the incident beam direction, giving rise to magnetic contrast in the imaging plane. In Fig. 1b, stripe domain structures with alternating magnetization aligned along the easy axis inside domain are separated by Bloch-type domain walls with local magnetization perpendicular to the easy axis. Lorentz TEM (LTEM) contrast is depicted for two orthogonal imaging directions (i.e., projections on the *bc* or *ab* planes). For the *bc* plane projection, magnetic contrast is induced by magnetization inside domain, resulting in either bright or dark bands centered at domain walls. In the case of the *ab* plane projection, as all magnetic moments inside domains are parallel to the imaging direction, the LTEM contrast is only generated by domain walls, forming either bright-dark or dark-bright bands (bottom in Fig. 1b). Here, the magnetization directions in domain walls are assumed to be also alternating across each domain.

In Figure 2, we show ground-state magnetic domain structures at low temperature in CGT TEM sample prepared by FIB along the *bc* plane projection. We fabricated the TEM sample with four sections in a range of discrete TEM foil thicknesses (91 ~ 301 nm), estimated by their electron-energy-loss spectra (EELS). The magnetic contrast shows up, as shown in Fig. 2b, when the sample was cooled below 60 K, a few K below the reported $T_c$ for bulk crystals[11]. A zoom in of this observed contrast shown in Fig. 2c is consistent with the stripe domain structure schematic depicted in Fig. 1b along the *bc* plane projection and demonstrates the strong magneto-crystalline anisotropy of CGT that persists despite additional shape anisotropy terms arising in the cross-sectioned TEM sample. The domain width varies with the TEM foil thickness (see Fig. 2b, also Fig. S2 in the Supplementary Information), although a more systematic study is needed to understand the exact thickness-dependence of stripe domains. The phase transition temperature ($T_c$) and critical magnetic field ($H_c$) for poled ferromagnetic spin states are also TEM foil thickness dependent (Fig. 2d), which indicates a tangible effects of boundary conditions on stripe domain structures. Both $T_c$ and $H_c$ decrease with decreasing TEM foil thickness. For the 91 nm thick section, the $T_c$ was 57 K, which is considerably lower than the bulk value of 65 K. This fact is consistent with the drastically decreased $T_c \sim 35$ K reported for double layer CGT film[11]. It should be noted that, due to discrete changes in thickness across each section, the leakage of stray demagnetizing field may influence the stability and width of domains in its vicinity.

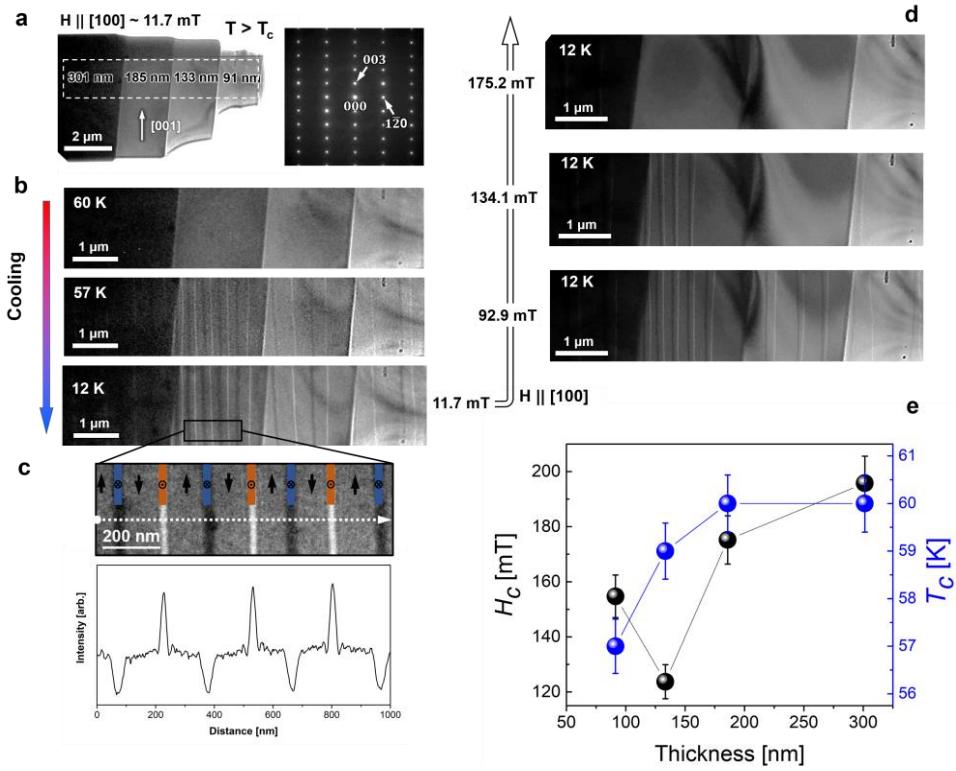

**Fig. 2| Stripe domains in ground state Cr$_2$Ge$_2$Te$_6$. a**, LTEM image and electron diffraction patterns obtained at T ~ 60 K. Local thicknesses estimated by EELS spectra are shown. Samples are cooled under residual magnetic field 11.7 mT. **b**, Temperature-series of LTEM Images upon cooling. Magnetic-field dependence of stripe domains at 12 K **c**, Enlarged LTEM image showing the stripe domains separated with Bloch walls with alternating out-of-plane spins, as schematically shown. A line profile across the stripe domains is consistent with the alternating black and white contrast in the *bc* plane projection in Fig. 1b. **d**, The magnetic-field dependence of the stripe domains at 12 K. All LTEM images were taken with defocus value ~ 200 μm. **e**, The thickness-dependent critical magnetic field ($H_c$) and the paramagnetic-to-ferromagnetic phase transition temperature ($T_c$). Both $H_c$ and $T_c$ were determined from LTEM images where magnetic contrast completely disappears for each thickness sections.

We investigated low-$T$ magnetic spin textures on *ab* plane in the sample that is mechanically exfoliated on polymer poly-dimethoxy silane (PDMS) stamps and dry transferred with micron level precision onto silicon nitride TEM grid with 10 μm x 10 μm observation window. Before exfoliation, the silicon nitride membrane (20 nm thick) on the observation window was removed to enhance Lorentz microscopy contrast. The narrow observation window keeps the exfoliated CGT sample flat to avoid an in-plane component of external magnetic field due to tilting between applied magnetic field along the imaging direction and the *c*-axis as well as reduces strain from bending of the exfoliated flake. It is known that magnetic spin textures in bubble materials are strongly dependent on magnetic field history. We have investigated spin textures in both residual-field (lowest magnetic field in TEM, 11.7 mT) and intermediate-field (with an excitation of objective lens, 50 mT) cooling cases, as shown in Fig. 3. For residual-field cooling case, the stripe domains with alternating spin aligned along the easy axis evolve and some bubble domains as well, as shown in Fig. 3a. With increasing external magnetic field, the stripe domain become narrow and turn into bubble shapes, that are sparsely distributed. In 50 mT field cooled sample, densely packed bubble domains were observed (Fig. 3e), similar to hexagonal bubble lattices reported in other bubble systems such as orthoferrites, hexagonal ferrites, and uniaxial garnets[20]. With increasing magnetic field, the bubble size decreases and finally disappears in the ferromagnetically poled state. The detailed spin arrangements around stripe domains and bubbles are revealed in Figs. 3i-j by the transport-intensity-equation (TIE) method. The stripe domains with alternating inwards and outwards spins are separated by Bloch domain walls that also show alternating in-plane spins. These type-I bubbles are topologically identical to that of Bloch skyrmions induced by DMI with both possessing topological charge of +/- 1.

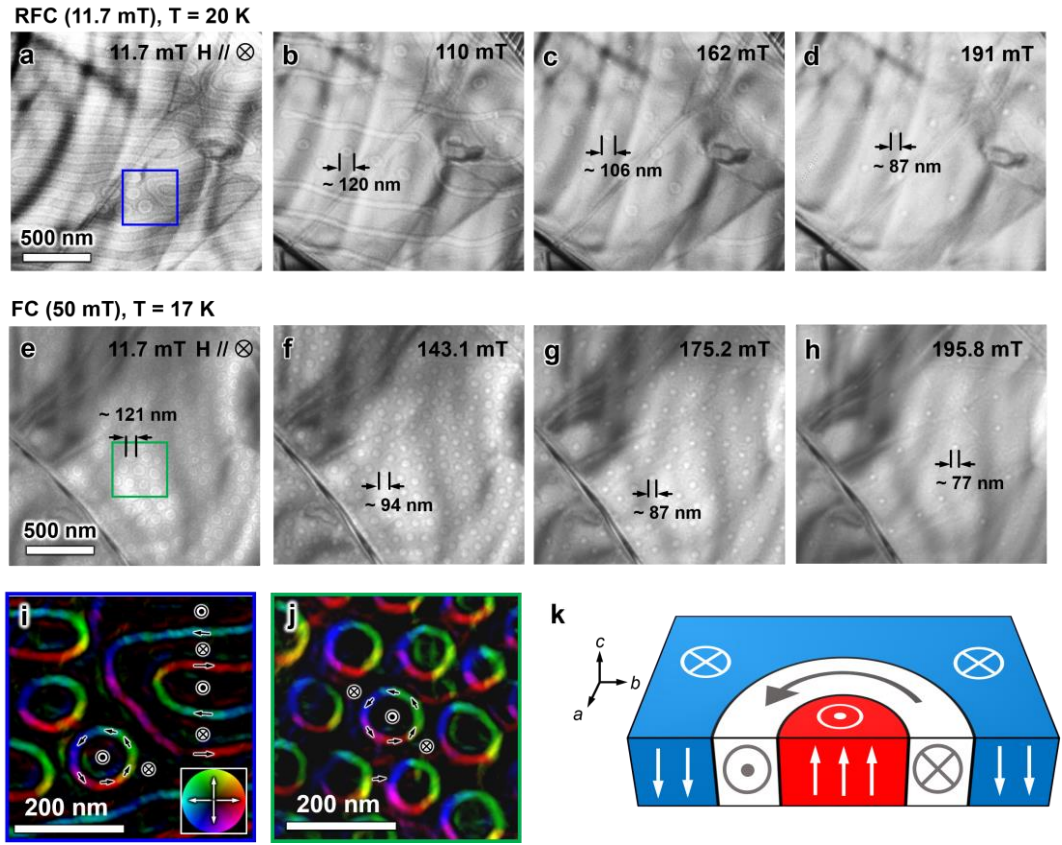

**Fig. 3| Stripe-Skyrmionic bubble spin transition induced by magnetic field in an exfoliated van der Waals $Cr_2Ge_2Te_6$. a-d**, LTEM images showing magnetic-field-driven transitions from stripes (**a-b**) to skyrmionic bubbles (**c-d**) in residual-field (11.7 mT) cooled sample. External magnetic field (*H*) was applied along the imaging direction (i.e., along the *c*-axis, pointing into the *ab* plane of exfoliated flakes). **e-h**, Magnetic field dependence of bubble lattices in 50 mT field cooled sample. All LTEM images were taken with defocus value ~ 200 μm. **i-j**, Magnetization maps obtained by transport-of-intensity equation (TIE) analysis of focal series of LTEM data from the areas indicated with blue (**i**) and green (**j**) boxes in **a** and **e**, respectively. The skyrmionic bubble sizes are inversely scaled with external magnetic field (**f**) before annihilation in ferroelectrically poled state above *H* > 200 mT. **k**, Schematic of the skyrmionic bubble spin

However, it should be noted that spatial spin arrangement differs in type-I bubbles from that in Bloch skyrmions in chiral magnets; magnetic spins are rapidly changing across domain walls and otherwise almost uniform for bubbles while the spatial variation of spin is almost uniform from the center to its surrounding in the case of Bloch skyrmions. In addition, unlike Bloch skyrmions where their size is fixed by the exchange interaction and DM interaction, the size of bubbles can be further tuned by external magnetic field, as shown in Fig 3a-h. This observation presents strong evidence that the bubble shaped spin-textures observed in our CGT samples are not Bloch skyrmions.

Interestingly, we observe bubbles of a single chirality (with the same sense of in-plane spin rotation) in contrast to conventional bubble lattices where random chirality was reported[22,32]. For applications in vdW 2D heterostructures based memory or logic devices, the presence of a single chirality bubble lattices would be preferable as it allows for their predictive displacement upon application of spin-polarized currents.

The single-chiral bubble formations in CGT indicates that the stripe domains are bounded by alternating Bloch domain walls. During dynamic transition from stripe domains to bubbles, we found that pairs of Bloch lines can be nucleated and annihilated in stripe domains, leading to spin rotation within Bloch walls. These Bloch lines are called π-Bloch lines because spin rotate +/-180° across each line[33]. Thus, these Bloch lines are also considered as topological defects with +/- 1/2 charges where integer topological charges associated with each stripes or bubbles are preserved during their pair formation and annihilation process. According to prior reports on conventional bubble domains, these Bloch lines are not energetically expensive and may effectively reduce stray fields compared to the simple Bloch domain walls without Bloch lines[21]. In Fig. 4a, we observe two π-Bloch lines with opposite sense of rotation are formed in otherwise two pairs of Bloch walls in narrow stripe domains. Interestingly, these π-Bloch lines are the point where two head-to-head or tail-to-tail spins are met. In Fig. 4d, two sets of π-Bloch lines are formed at the charged segments of stripe domains; as two π-Bloch lines are associated with each point, two pairs of head-to-head or tail-to-tail spin arrangements are formed. Here, we define the narrow stripe domains bounded by two antiparallel Bloch walls as type-I stripe domains and the other stripe domains with two parallel Bloch walls as type-II. We also observe that type-II stripe domains can be transformed into type-I through pair annihilation of Bloch lines with opposite topological charges.

The effects of Bloch lines on topological structures of bubble domains are elaborated in Figure 4 e-g. With persisting Bloch-line pairs, each type of stripe domains continuously turn into bubbles with the same type since the topological charges are preserved. In Fig. 4e, two type-I stripe domains and one type-II (middle) stripe domain under 70 mT magnetic field are shown. With a slight increase in magnetic field to 90 mT, the type-II stripe domain shrinks significantly while two type-I domains remain unaffected. The type-II stripe turns into type-I bubble without Bloch lines, as shown in Fig. 4f. This indicates the two π-Bloch lines at the end of the type-II stripe were pair-annihilated when they are brought into proximity due to increasing external magnetic field. This pair-annihilation process "unwinds" the topological charges that cancel, transforming the type-II stripe into type-I bubble. This unwinding process explains the dominance of single-chiral type-I bubbles induced by magnetic field with/without field cooling process. In fact, this homochiral type-I bubble would be superior to random chirality type-I bubbles as it provides better control over displacements for real device applications.

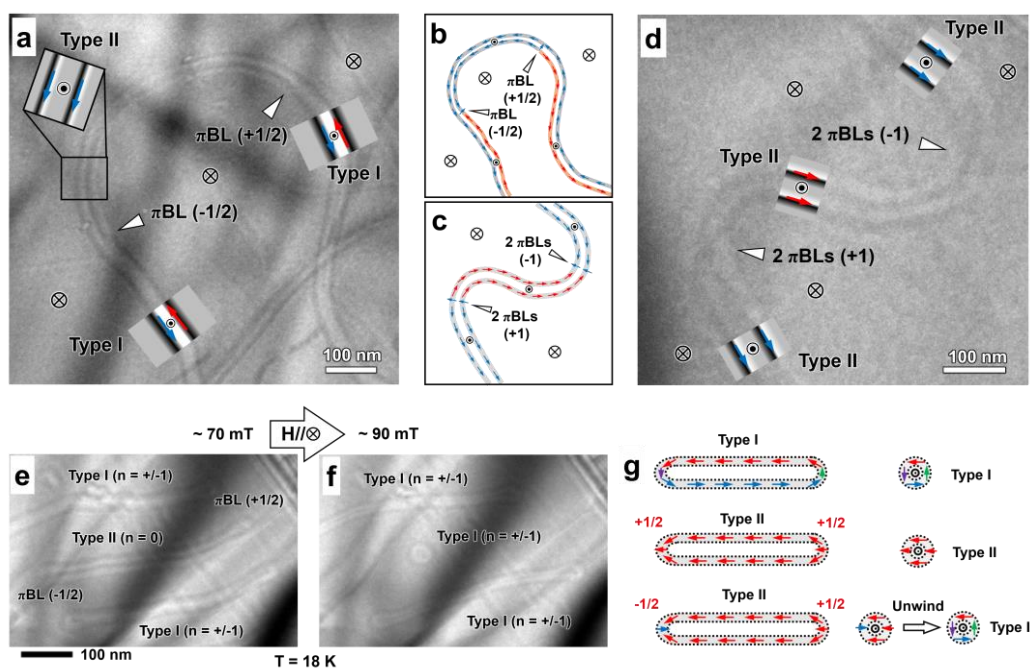

**Fig. 4| Bloch-line pairs and their unwinding under external magnetic fields. a-b**, LTEM images showing type-I and type-II stripe domains bounded by a pair of π BLs and its schematic. **c-d**, LTEM image showing type-II stripe domain with a pair of 2π BLs and its schematic. **e-f**, LTEM images showing transformation of type-II stripe into type-I bubble under a slight increase in magnetic field. **g**, schematics of stripe-bubble transformation process including "unwinding", a pair annihilation of π BLs.

In conclusion, topologically nontrivial spin textures, skyrmionic bubbles and Bloch lines, are directly observed for the first time in exfoliated 2D ferromagnets. Topological transformation between type-I and type-II stripes through pair formation and annihilation of Bloch lines demonstrate rich spin textures available in 2D ferromagnets. Topological spin textures observed in 2D vdW system provide new platform to realize novel quantum states utilizing magnetic-proximity effects.

**Methods**

$Cr_2Ge_2Te_6$ synthesis: Single crystal $Cr_2Ge_2Te_6$ was grown by self-flux technique from a mixture of pure Cr power, Ge and Te pieces in a molar ratio of 1 : 2 : 6. The mixture was heated in an evacuated quartz tube to 1100 ºC for over 20 hours and then cooled to 700 ºC with 1 ºC/h cooling rate. The detailed synthesis information can be found in Ref. 30.

TEM analysis: Cross-section TEM samples were prepared by focused-ion beam (FIB) technique with 2 keV $Ga^+$ to reduce ion-beam-induced damage. A JEOL ARM 200CF equipped with cold field emission gun and double-spherical aberration-correctors at the Brookhaven National Laboratory was used for HAADF STEM imaging and LTEM. The range of collection angle for HAADF STEM was 68 to 280 mrad. A double-tilt liquid helium cooling holder (HCTDT 3010, Gatan, Inc.) was used for low temperature experiments. The defocus value for Lorentz imaging was about ± 200 μm. For EELS, Gatan Quantum ES spectrometer was used with 0.5 eV/ch dispersion and 1.5 eV energy resolution. The convergent and collection semi-angles were, respectively, ~ 10 and ~ 5 mrad. In order to obtain the absolute thicknesses of TEM samples, the mean free path for inelastic scattering of CGT was estimated on the basis of a dipole formula with an effective atomic number based on the Lenz model (see page 295-301 in Ref. 34 for more details)[34].

**Acknowledgement**

This work is supported by the US DOE Basic Energy Sciences, Materials Sciences and Engineering Division under Contract No. DE-SC0012704. D.J. and H.Z. acknowledge primary support from Northrop Grumman Corporation and partial support from National Science Foundation (DMR-1905853), University of Pennsylvania Materials Research Science and Engineering Center (MRSEC) (DMR-1720530) and contract number W911NF-19-1-0109 from the Army Research Office. Research carried out, in part, at the Center for Functional Nanomaterials, Brookhaven National Laboratory, which is supported by the US Department of Energy, Office of Basic Energy Sciences, under Contract No. DE-AC02-98CH10886.

**Conflict of Interest:** The authors acknowledge no conflict of interests.

**Supporting Information**

The Supporting Information is available for free of charge on the ACS Publications website.

EELS spectra for thickness measurements, Thickness-dependent domain width, and EDS chemical composition analysis.

**Table of Content**

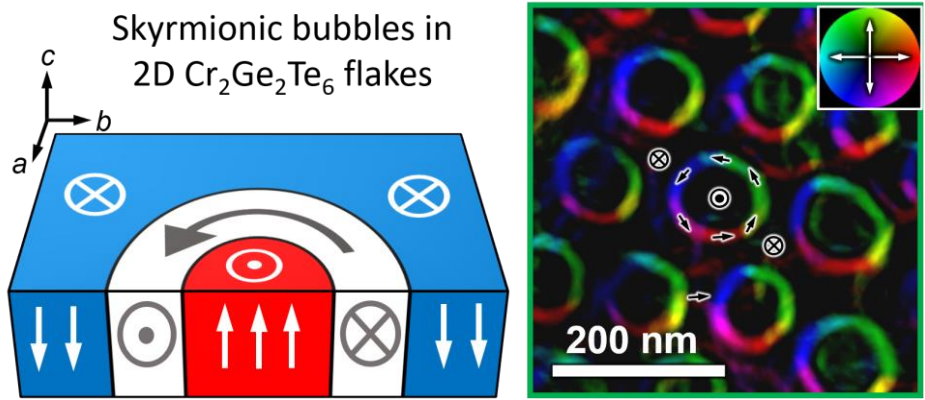